\documentclass[12pt]{article}
\usepackage{times} \usepackage{cite} \usepackage{amssymb}
\usepackage{pifont} \usepackage{amsmath}
\usepackage{amscd}
\numberwithin{equation}{section}
\usepackage{ifthen}\usepackage{calc} 
\newcounter{Pol}\newcounter{Dpol}\newcounter{Spol}\newcounter{Mpol}
\newcommand\varw{w}
\newcommand\polsubc{}\newcommand\varzw{w}
\newcommand\polynomials[2]{
\setcounter{Mpol}{#1}              
\setcounter{Pol}{1}
\setcounter{Dpol}{1}
\setcounter{Spol}{1}               
\whiledo{\value{Pol} < #2}{
\setcounter{Dpol}{\value{Dpol}*10}
\stepcounter{Pol}
}
\setcounter{Pol}{0}
\whiledo{\value{Pol} < #2}{
\setcounter{Spol}{\value{Mpol} / \value{Dpol}}
\ifx\varzw\varw\renewcommand\varzw{z}
\else\renewcommand\varzw{w}\fi
\renewcommand\polsubc{{\varzw}_{\theSpol}}
{\polsubc}
\addtocounter{Mpol}{-\value{Spol}*\value{Dpol}}
\setcounter{Dpol}{\value{Dpol}/10}
\stepcounter{Pol}
}
}
\newcommand\zpol[2]{\renewcommand\varzw{w}\polynomials{#1}{#2}}
\newcommand\wpol[2]{\renewcommand\varzw{z}\polynomials{#1}{#2}}

\newcommand\preprint[1]{\vspace{-1in}\vtop{\null\hfill
\parbox[t]{1.6in}{\small\sc #1\\\null}}
\vskip .5in\bigskip\normalfont}

\renewcommand\bar\overline
\newcommand\cf[4]{\bibitem{#1}{#2}.~{\it #3}\,;~{#4}.}

\def\coker#1{\qopname\relax o{coKer}~{#1}}

\def\C{{\mathbb{C}}}

\def\diag#1{\text{Diag}\{{#1}\}}

\newcommand\eg{{\slshape e.g.~}}

\def\eq#1{(\ref{#1})}

\def\FI{Fayet--Iliopoulos~}

\def\goth#1{{\mathfrak #1}}

\def\id#1{{\mathbb{I}}_{#1}}
\newcommand\ie{{\slshape i.e.~}}
\def\ker#1{\qopname\relax o{Ker}~{#1}}
\def\map{\longmapsto}

\def\pa{\partial}

\def\rt{\longrightarrow}

\def\tilde{\widetilde}
\def\tr#1{\qopname\relax o{Tr}~{#1}}
\def\T{{\mathbb{T}}}
\def\viz{{\slshape viz.~}}
\def\Z{{\mathbb{Z}}}
\def\w{\mathsf{w}}
\def\z{\mathsf{z}}
\def\be{\begin{equation}}
\def\ee{\end{equation}}
\newcommand\atmp[3]{Adv. Theor. Math. Phys. {\bf #1}~(#2)~#3}
\newcommand\alg[1]{alg-geom/{#1}}
\newcommand\cmp[3]{Comm. Math. Phys.{\bf #1}~({#2})~{#3}}

\newcommand\hepth[1]{hep-th/{#1}}

\newcommand\jgp[3]{J. Geom. Phys. {\bf {#1}}~({#2})~{#3}}
\newcommand\jhep[3]{JHEP {\bf {#1}}~({#2})~{#3}}
\newcommand\npb[3]{Nucl.~Phys. {\bf B{#1}}~({#2})~{#3}}

\newcommand\plb[3]{Phys. Lett. {\bf B{#1}}~({#2})~{#3}}
\newcommand\prd[3]{Phys. Rev. {\bf D{#1}}~({#2})~{#3}}

\begin{document}
\title{\preprint{rom2f--99/30\\mri-phy/P990927\\hep-th/9909107}
D-branes on Fourfolds with Discrete Torsion}
\author{}
\date{
Subir Mukhopadhyay
\thanks{E--mail:~ subirm@mri.ernet.in 
~~[Address after 24 Sept, 1999 : 
Fysikum, Stockholm university, Box 6730, S-113 46, Stockholm, Sweden] }
\\{\small{\sl Mehta Research Institute of
Mathematics and Mathematical Physics,\\ Chhatnag Road, Jhusi,
 Allahabad 211 019, India.}
}
\\{\sl\&}\\ Koushik Ray
\thanks{E--mail:~koushik@roma2.infn.it}
\\{\small
{\sl  Dipartimento di Fisica,\\ Universit{\`a} di Roma  ``Tor Vergata'', 
 INFN  --- Sezione di Roma  ``Tor Vergata'',\\
Via della Ricerca Scientifica, 1,
00173  Rome, Italy}
}}%
\maketitle
\thispagestyle{empty}
\vfil
\begin{abstract}
\noindent We study D1-branes on the 
fourfold $\C^4/(\Z_2\times\Z_2\times\Z_2)$, 
in the presence of discrete torsion. 
Discrete torsion is incorporated in the gauge theory of 
the D1-branes by considering a projective representation of the 
finite group $\Z_2\times\Z_2\times\Z_2$. The corresponding orbifold
is then deformed by perturbing the F-flatness condition of the 
gauge theory. The moduli space of the resulting gauge theory
retains a stable singularity of codimension three. 
\end{abstract}
\clearpage
\section{Introduction}
D-branes provide a geometric means to studying orbifold
singularities and their desingularisations, as the moduli space
of D-branes reproduces the space in which the D-branes are embedded into.
Within the scope of string theory, a generalisation of orbifold 
singularities, when possible, is to turn on a discrete 
torsion\cite{vafa,vafwit}. 
String theory on an orbifold of the $n$-dimensional complex space $\C^n$, 
\eg $\C^n/G$, where $G$ is a finite group,
admits discrete torsion if the second 
cohomology group of $G$, \viz $H^2(G,U(1))$, is non-trivial. 
At this point, let us recall that here we are considering D-branes
on non-compact spaces, which may serve as local models of  a compact 
target space of string  theory near a singularity. 
In the conformal field theoretic description of the string world-sheet, 
turning on a discrete torsion is tantamount to
assigning a non-vanishing weight or phase to the contribution to 
the string partition function arising in the twisted sector.

In studying a closed string theory on an orbifold in absence of discrete 
torsion, one can estimate the contributions to the partition function
arising in the untwisted and the twisted sectors of the theory, by 
implementing the quotient in the path integral of the theory. 
The resulting spectrum, which is the sum of all these contributions,
turns out to coincide with that of the theory on the corresponding
smooth manifold \cite{aspin}. 

The presence of discrete torsion alters the contribution
from the twisted sector of the theory. The resulting theory is still 
consistent as a string theory, but no more a string theory on a blown
up manifold. Indeed, the resulting theory might be a consistent string
theory on a singular target space \cite{vafwit}. Moreover, the modes
in the twisted sector correspond to (partial) deformation of 
the complex structure of the orbifold, not to the blowing up of 
the K\"ahler class. Now that the target space of string theory 
can be simulated as the moduli space of D-branes,
a natural question is whether one can incorporate 
discrete torsion in this picture. This has been answered in the 
affirmative in some examples \cite{dt1,dt2}. In this 
article we will find one more example of this kind.

In terms of the supersymmetric gauge theories used to describe 
the theory on the world-volume of  
D-branes, discrete torsion is incorporated in the  
action of the quotienting group on the position as well as
the gauge
degrees of freedom carried by the brane. In the presence of discrete 
torsion one is led to choose a \emph{projective representation}
of the group \cite{dt1,dt2}, to be contrasted with the 
\emph{linear representation} used when discrete torsion is absent.
The resolution of the orbifold in the absence of discrete
torsion is effected by adding a \FI term in the 
gauge theory, thereby perturbing the D-term of the gauge theory.
In the presence of discrete torsion, the moduli 
for the deformation of the singularity are purveyed by 
parameters appearing in the perturbation of the F-term of the 
theory. In either case, the choice of the perturbations of the F- and 
D-terms are guided by the twisted sector of closed string theory
in the presence and absence of discrete torsion, respectively
\cite{dm,dgm,dt1,dt2,jm}. 

D-branes on the three-dimensional orbifold
$\C^3/(\Z_2\times\Z_2)$ have been studied 
in the absence of discrete torsion \cite{brg,subir,morples}, and in its 
presence as well\cite{dt1,dt2,gomis}. 
In the absence of discrete torsion, the moduli space of a D-brane
on $\C^3/(\Z_2\times\Z_2)$ is a blown down conifold. Adding \FI terms 
in the gauge theory corresponds to partial \cite{morples} or complete 
\cite{brg,subir} resolution of the singularity, depending on the 
non-vanishing combinations of the \FI parameters.
However, the scenario is rather different in the presence of discrete
torsion \cite{dt1,dt2}. In this case, 
the twisted sector of string theory provides 
modes not to resolve the singularity, but to deform it. Yet,
the moduli space turns out to contain 
a stable double-point singularity or node, while the 
codimension-two singularities are deformed away by these modes.
This signals a deficiency of certain modes in the 
twisted sector of the closed string theory corroborating with 
earlier findings \cite{vafwit}.

In the present article we will concern ourselves with analysing a
D1-brane or D-string on $\C^4/(\Z_2\times\Z_2\times\Z_2)$
in the presence of discrete torsion. D1-branes on this space 
without discrete torsion has been studied earlier\cite{ahn}. 
In the absence of discrete torsion, the analysis of the moduli space of 
D1-branes on $\C^4/(\Z_2\times\Z_2\times\Z_2)$ parallels the 
analysis of D3-branes on $\C^3/(\Z_2\times\Z_2)$. The 
theory of D1-branes on the singular orbifold is an 
$\mathsf{N}=(0,2)$ super Yang-Mills
theory in two dimensions. In the absence of discrete torsion
the singularity in the moduli space is resolved by introducing 
\FI terms in the action of the super Yang-Mills theory.
The moduli space with perturbed D-flatness conditions, 
implementing the resolution of the orbifold, can be 
studied using the paraphernalia of toric geometry, where the monomials
of the toric description are provided by the unperturbed F-flatness
condition of the theory. 
In the presence of discrete torsion, however, it is the F-term 
of the $(0,2)$ theory that admits a perturbation, involving 
six parameters. The D-flatness conditions, however, remain unaltered 
with respect to the theory on the singular orbifold.
This corresponds to a deformation of the singularity, 
rather than its resolution. The perturbation of the F-flatness conditions
prevents employment of toric geometry 
in the description of the deformed variety. 
One is led to consider the gauge-invariant quantities to furnish a 
description of the deformed variety.
It is found that, after desingularisation by F-term perturbations,
the variety describing the moduli space of the gauge theory 
retains a stable singularity of codimension three (line singularity), 
in the same manner as  its
three-dimensional counterpart retains a conifold singularity
after deformations\cite{dt1,dt2}. 

The plan of the article is as follows. 
We recount some features of the projective representations 
of $\Z_2\times\Z_2\times\Z_2$ 
in \S\ref{proj}. In \S\ref{closed} we study  
the twisted sector of closed string theory on 
$\C^4/(\Z_2\times\Z_2\times\Z_2)$ in the presence of 
discrete torsion. This analysis enables us to determine the number 
of perturbation parameters allowed in the gauge theory. The low 
energy effective gauge theory of the D-brane on the orbifold is discussed 
in the \S\ref{gaugesec}. 
In \S\ref{secdeform} we consider the 
vacua of the resulting gauge 
theory and find out the corresponding moduli 
spaces both with and without deformations, before 
concluding in \S\ref{conclud}. 

\vspace{1cm}  

\par\noindent\underline{\sc Notations and conventions:}
Unless explicitly declared otherwise, we follow the following conventions
in notation and terminology in the sequel.
\begin{itemize}
\item The terms \emph{resolution} (or \emph{blow up}) and 
\emph{deformation} of singularities 
are used in the usual senses. The term \emph{desingularisation}
is used generally to mean either of these two, therby encompassing 
partial removal of singularities.
\item  For subscripts,
uppercase letters from the middle of the alphabet, \eg $I$,$J$, $K$, etc.
assume values in $\{1,2,3,4\}$, while the corresponding lower-case letters,
namely, $i$, $j$, $k$ etc., assume values in $\{1,2,3\}$.
\item The Pauli matrices $\sigma_I$ are chosen to be the following:
\begin{equation}\nonumber
\sigma_1 = \begin{pmatrix}
0&1\\1&0
\end{pmatrix}\quad
\sigma_2 = \begin{pmatrix}
0&-i\\i&0
\end{pmatrix}\quad
\sigma_3 = \begin{pmatrix}
1&0\\0&{-1}
\end{pmatrix}\quad
\sigma_4 = \begin{pmatrix}
1&0\\0&1
\end{pmatrix}
\end{equation}
\item No sum is intended on repeated indices.
\end{itemize}
\section{Projective representation \& discrete torsion}\label{proj}
In considering desingularisations of orbifolds of the type $\C^4/G$, 
discrete torsion corresponds to non-trivial elements of the 
second cohomology group of the finite group $G$.\footnote{See, however, 
\cite{aspmor} for more rigorous considerations on discrete torsion.} 
It is incorporated in the theory through the action of the group $G$ on the 
Chan-Paton degrees of freedom by using an adjoint action by 
a projective representation of $G$.
In this section we will collect some facts \cite{kap} about 
the projective representation of $G$, relevant for the present article. 

Given a finite group $G$,  a mapping 
$\rho : G \rt GL(n, \C)$ is called a \emph{projective 
$\alpha$-representation} of $G$ (over the field $\C$), provided there
exists a mapping $\alpha : G\times G \rt U(1)$, such that 
\begin{enumerate}
\item $\rho (g)\rho(g') = \alpha(g,g')\rho(g g')$,
\item $\rho(\mathbf{1}) = \id{n}$,
\end{enumerate}
for all elements $g, g' \in G$, where $\mathbf{1}$ denotes the 
identity element of $G$ and $\id{n}$ denotes the $n\times n$ identity 
matrix in $GL(n, \C)$. 
Let us note that one can define 
a projective representation over more general fields \cite{kap}.
Here we are referring to the projective matrix 
representation over the field of complex numbers as the projective 
representation. It can be shown that $\alpha$ is a $U(1)$-valued 
two-cocycle of the second cohomology group $H^2(G, U(1))$ of 
the finite group $G$. For our purposes $\alpha$ will be a complex number 
with unit modulus. 
\footnote{Usually the range of the map $\alpha$ is 
taken to be the multiplicative group $\C^{\star}$ of the field $\C$ of 
complex numbers. Here we will only consider the map $\alpha$ with unit
modulus. So, we have taken the range to be $U(1)$.} 

Generally, the second cohomology group $H^2(G,U(1))$ 
of a direct-product group of the form 
$G= {\displaystyle\bigotimes_1^m \Z_n}$, is isomorphic to 
${\displaystyle\bigotimes_{1}^{m(m-1)/2}\Z_n}$. Let $\goth{g}_i$  
denote the generator of the $i$-th $\Z_n$ factor appearing in $G$, \ie
$G =\displaystyle\bigotimes_1^m \langle \goth{g}_i\rangle$. 
Let $g_i$, $i=1,2,\cdots , m$, 
denote the generators of $G$. Let us also define
\begin{eqnarray}\label{mubeta} 
\mu(i) = \prod_{a=1}^{n-1}
\alpha(g_i^a,g_i) \quad\text{and}\quad \beta(i,j) 
= \alpha(g_i,g_j)\alpha(g_j,g_i)^{-1}, \quad i,j = 1,2,\cdots , m. 
\end{eqnarray} 
We may set $\beta$ to be an $n$-th root of unity and $\mu =1$,
by replacing $\alpha$, if necessary, by a cohomologous cocycle. 
The corresponding projective $\alpha$-representation of $G$ is 
given in \cite{kap} for special values of $\beta$ . 
For our purposes it suffices to 
quote the results for the special case with $m=3$ and $n=2$.

Thus, we will consider the projective representations of 
the group $G = \Z_2\times\Z_2\times\Z_2$. Let us assume that $G$ has the 
following action on the four coordinates $Z_1$,$Z_2$,$Z_3$,$Z_4$ of $\C^4$:
\begin{equation}\label{sptimeact}
\begin{split}
g_1 : (Z_1, Z_2, Z_3, Z_4) &\map
(-Z_1, -Z_2, Z_3, Z_4), \\
g_2 : (Z_1, Z_2, Z_3, Z_4) &\map 
(-Z_1, Z_2, -Z_3, Z_4),  \\
g_3 : (Z_1, Z_2, Z_3, Z_4) &\map 
(-Z_1, Z_2, Z_3, -Z_4).
\end{split}
\end{equation}
Let $\goth{g}_1$, $\goth{g}_2$ and 
$\goth{g}_3$ denote the three generators of the three  $\Z_2$ factors.
A generic element of $G$ can be written as 
$g = \goth{g}_1^a \goth{g}_2^b \goth{g}_3^c$. We will 
denote this element by the symbol $(abc)$. 
Choosing  the action of each of the $\Z_2$ factors to be a change of sign 
of $Z_1$ and one more out of $Z_2$, $Z_3$ and $Z_4$, we can write $g_i$ from 
\eq{sptimeact} as
\begin{equation}\label{genGex}
\begin{split}
g_1 =\goth{g}_1 &= (100),\\
g_2 =\goth{g}_2 &= (010),\\
g_3 =\goth{g}_3 &= (001).
\end{split}
\end{equation}
The second 
cohomology group $H^2(G,U(1))$ of $G$ is isomorphic to
$\Z_2\times\Z_2\times\Z_2$ \cite{kap}. The three generators 
of the latter may be taken to be 
\begin{equation}
\alpha_1 ((abc), (a'b'c')) = i^{(ab'-ba')},\quad
\alpha_2 ((abc), (a'b'c')) = i^{(bc'-cb')},\quad
\alpha_3 ((abc), (a'b'c')) = i^{(ca'-ac')}.
\end{equation}

Let us note that $\alpha_i((abc),(abc)) = 1$ for $i=1$, 2, 3. 
Hence, by \eq{mubeta}, we have $\mu (i) =1$, for $i=1,2,3$.
In what follows we will consider the element
$\alpha=\alpha_1\alpha_2\alpha_3$. Thus, we have, for all $g, g'\in G$,
\begin{equation}\label{eqnalpha}
\begin{split}
\alpha(g,g') &= i\quad\text{if~} g\neq g',\\
&= 1\quad\text{if~} g=g',
\end{split}
\end{equation}
and consequently, $\mu(i)=1$ for $i=1,2,3$, and $\beta = -1$. 
There are two irreducible $\alpha$-representations
of $G$, which are not linearly equivalent. These are given by\cite{kap}
\begin{equation}\label{REP} 
\rho(g_i) = \pm \sigma_i.
\end{equation}
The discrete torsion appearing in the path integral is determined 
by the choice of the two-cocycle. For example, 
for the above choice of the two-cocycle, namely $\alpha = 
\alpha_1\alpha_2\alpha_3$, 
the discrete torsion is given by \cite{dt1,dt2} 
\begin{equation}\label{epsil}
\begin{split}
\varepsilon ((abc),(a'b'c')) 
&= \Big(\alpha ((abc),(a'b'c'))\Big)^2
,\qquad (abc)\neq (a'b'c') \\
&= (-1)^{ab'-ba'+bc'-cb'+ca'-ac'},
\end{split}
\end{equation}
which is the form used in \cite{acharya}.
Each  given value of the discrete torsion corresponds to 
a variety with a different 
topology and leads to different types of gauge 
theories. These different varieties are 
related by mirror symmetry \cite{acharya}.

Let us point out that the discrete torsion $\alpha$  used in \eq{eqnalpha}
is by no means more general than any other two-cocycle. This can 
be interpreted as a phase between two of the $\Z_2$ factors in $G$
by a change of basis. This fact will be 
reflected in the moduli space of the brane in that, the maximally
deformed moduli space retains a singular line, unlike the
point in \cite{dt1,dt2}.
\section{String(ent) restrictions on deformations}\label{closed}
It has been known that string theory can be defined 
on certain kinds of singular spaces, especially on orbifolds.
In sigma-model compactification on orbifolds,
the spectrum of string theory receives 
contributions from the twisted sectors, thereby rendering string  theory 
well-defined on such spaces. In considering D-branes on orbifolds,
the orbifold is realised as the moduli space of the gauge theory  on the 
world-volume of the brane. Resolution or deformation of the quotient 
singularity is effected by perturbing the gauge theory.
However, compatibility of the desingularised 
D-brane moduli space with string theory imposes stringent restrictions
on such extra terms. In this section, we discuss these restrictions.

The four-dimensional orbifold $\C^4/(\Z_2\times\Z_2\times\Z_2)$ 
in the blown down limit may be defined as an affine variety embedded 
in $\C^5$ by the polynomial equation $\mathcal F(x,y,z,w,t) = 0$, where
the polynomial is
$\mathcal F(x,y,z,w,t) = xyzw -t^2$, and $x$, $y$, $z$, $w$, $t$ 
are the coordinates of $\C^5$.
The $\Z_2\times\Z_2\times\Z_2$ symmetry can be made conspicuous 
by expressing the coordinates of $\C^5$ in terms of the 
affine coordinates of the covering space of the variety, namely, $\C^4$.
Explicitly, $x=u_1^2, y=u_2^2, z=u_3^2, w=u_4^2, t=u_1u_2u_3u_4$, with
$\{u_1,u_2,u_3,u_4\}\in\C^4$. The polynomial $\mathcal F$ remains invariant 
under the three independent transformations, which change the signs
of, say $u_1$, together with one out of $\{u_I| I=2,3,4\}$ in turn.

Let us briefly recount the algebraic deformations of the equation 
$\mathcal F =0$. The possible deformations are 
given by the ring of polynomials 
$Q=\C[x,y,z,w,t]/\langle\pa\mathcal F\rangle$, where 
$\partial\mathcal F$ 
stands for the set of the partial derivatives of $\mathcal F$ with respect 
to each of the arguments, \ie 
\[\pa\mathcal F = \{\pa\mathcal F/\pa x, 
\pa\mathcal F/\pa y, 
\pa\mathcal F/\pa z, 
\pa\mathcal F/\pa w, 
\pa\mathcal F/\pa t\},\] and $\langle\cdots\rangle$ denotes 
the ideal generated by $\cdots$. 
Thus, we have 
\begin{equation}
\begin{split}
\langle\pa\mathcal F\rangle &= \langle xyz, yzw, xyw, xzw, t\rangle ,\\
Q &= \langle 1, x^a, y^b, z^c, w^d, xy, xz, xw, yz, yw, zw\rangle ,
\end{split}
\end{equation}
where $a,b,c,d$ are arbitrary integers.
Among the generators of $Q$, terms such as $xy$ deform the six fixed 
planes such as $zw=0$, terms like $x$ deform the four fixed lines 
which correspond to codimension-three  singularities
and finally 1 deforms the double-point singularity with codimension four
at the origin, $t^2 + x^2 +y^2 +z^2 +w^2 = 0$.

However, as mentioned before, 
a physical theory, as the one we are considering,
is not necessarily potent enough to contain all the deformations
that are mathematically admissible. 
Within the scope of our discussion with D1-branes,
the four-dimensional orbifold is realised as the moduli space 
of an ${\mathsf N} = (0,2)$ super-Yang-Mills theory and its deformations
are subject to consistency requirements imposed by string theory. 
Considering branes in the closed or type-II string  theory, these 
consistency conditions are determined by the \emph{twisted sector}
of the theory on the orbifold. Of the generators
of the ring $Q$, only those deformations are physically allowed, that 
correspond to marginal operators in the closed string theory on the 
orbifold. The marginal operators are related by supersymmetry to the 
Ramond-Ramond (RR) ground states of the string theory. The latter, in turn,
are determined by the cohomology of the smooth target space that limits 
to the orbifold under consideration. To be more explicit, let 
$X$ be a manifold, and let $X'=X/G$ be an orbifold, where $G$ 
is a finite group of order $|G|$. Let $\tilde{X}$ be a desingularisation
of the orbifold $X'$. 
\begin{eqnarray}
\begin{CD}
&&&X\\
&&&@VGVV\\
\tilde{X}&@>>\text{desingularisation}>X'&=X/G
\end{CD}
\end{eqnarray}
Considering  string theory on the orbifold $X'$, the above-mentioned 
computation of RR ground states yields the cohomology 
$H^{\star}(\tilde{X})$ of $\tilde{X}$.

The desingularisation $\tilde{X}\rt X'$ can be effected in two different ways.
One way is to \emph{blow up} the singularity at the origin. 
This corresponds to  
turning on a \FI term in the ${\mathsf N}= (0,2)$ gauge theory,
and thereby perturbing the D-term of the gauge theory 
\cite{ahn}.  The other way is to \emph{deform} the singularity, discussed
above.  
This corresponds to perturbing the F-flatness conditions in the 
gauge theory and is relevant for us in considering orbifolds 
with discrete
torsion. At any rate, in order to count the physically admissible 
perturbation modes in the gauge theory
we need to consider closed strings on the orbifold 
and evaluate the cohomology. 

Let us compute the cohomology of the space $\tilde{X}$ for 
this case, following \cite{vafwit,dt2,zas,acharya}. 
The general strategy is as follows. Given an element $g$ of the 
group $G$, we first find out the subset of $X$ that is fixed 
under $g$. Let us denote this subset by $X_g$. The subset 
${X}_g$ is endowed with $(p,q)$-forms, denoted $\omega_g^{p,q}$.
Let $\Omega^{p,q}_g$ denote  the set of $(p,q)$-forms on ${X}_g$
and let $\tilde{\Omega}^{p,q}$ denote the set of all $(p,q)$-forms 
on $\tilde{X}$. 
The $(p,q)$-forms $\omega_g^{p,q}\in\Omega_g^{p,q}$, which are invariant 
under the group $G$, that is, which satisfy 
\begin{equation}\label{GinvDT}
\varepsilon (h,g) R(h)\omega_g^{p,q} = \omega_g^{p,q} , 
\qquad\omega_g^{p,q}\in \Omega^{p,q}_g,\quad h\in G, 
\end{equation}
contribute to $\tilde{\Omega}^{p+s,q+s}$, where
$s$ is the age of $g\in G$, defined by 
$s=\sum_{I=1}^{4} \vartheta_I$,
if $g:~Z_I \map e^{2\pi i\vartheta_I} Z_I$, where $Z_I$, $I=1,2,3,4$ 
are the coordinates of $X$.
Here $R(h)$ denotes some representation of the element $h\in G$ and 
$\varepsilon$ is the discrete torsion defined in 
\eq{epsil}.
The cohomology in absence of discrete torsion 
can be obtained by setting $\varepsilon =1$.

For the case at hand, the different elements of 
the group $G=\Z_2\times\Z_2\times\Z_2$ 
fixes three kinds of subsets in $X=\C^4$ --- contributing 
$h^{pq}$ $(p,q)$-forms to 
$\tilde{\Omega}^{p,q}$.
First, the identity of $G$ fixes
$\C^4$ itself, that is, ${X}_{(000)} = \C^4$.
We shall refer to the corresponding contribution to the cohomology as the 
contribution from the untwisted sector.
The $G$-invariant forms are 1, $dZ_I\wedge d\bar Z_I$ 
and $dZ_1\wedge dZ_2\wedge dZ_3\wedge dZ_4$ 
and some of the $dZ$'s replaced by their complex 
conjugates. The contribution to the cohomology $H^{\star}(\tilde{X})$ is 
summarised in the following Hodge diamond: 
\begin{eqnarray}
\begin{matrix}
&&&&1&&&&\\
&&&0&&0&&&\\
&&0&&4&&0&&\\
&0&&0&&0&&0&\\
1&&4&&12&&4&&1\\
&0&&0&&0&&0&\\
&&0&&4&&0&&\\
&&&0&&0&&&\\
&&&&1&&&&\\
\end{matrix}
\end{eqnarray}

These elements of $H^{\star}(\tilde{X})$ need be 
supplemented with the contribution from the 
twisted sector, that corresponds to the other two kinds of 
fixed sets. These are contributions from the $G$-invariant 
forms from the fixed subsets of $X$, under the non-trivial elements of $G$.
We will refer to these as the contribution from the twisted sector.
The element $(111)\in G$ fixes a point, ${X}_{(111)} = \{0\}$,  
the origin of $\C^4$, while each of the other six elements of $G$ leaves fixed 
a set $\C^2\subset X$. 

The contribution from the untwisted sector is the same both in the presence
and absence of discrete torsion. Contribution from the twisted 
sectors are different in the two cases. Let us consider both the cases in 
turn.
\begin{itemize}
\item {\sl Without discrete torsion:}\\
In the absence of discrete torsion the condition of $G$-invariance of the 
$(p,q)$-forms is given by \eq{GinvDT} with $\varepsilon =1$.

Only the $(0,0)$-form 1 is defined on ${X}_{(111)}$, which is a point.
This is obviously $G$-invariant.
The age of $(111)\in G$ is $s=2$. Thus, this form
contributes to $\tilde{\Omega}^{2,2}$,
with $h^{22}=1$. 

Each of the other six elements of $G$ has age $s=1$.
Each fixes a $\C^2\subset X$. For example, $(100)$ fixes the $\C^2$
coordinatised by $Z_3$ and $Z_4$.
The $G$-invariant forms on this $\C^2$
are 1, $dZ_i\wedge d\bar Z_i$, $i=3,4$ and $dZ_3\wedge 
dZ_4\wedge d\bar Z_3\wedge d\bar Z_4$. 
Taking into account the shift by the age of this element, the contribution 
to the Hodge numbers are:
$h^{11}=1$, $h^{22}=2$, $h^{33}=1$. 
Similar consideration with each of the other five elements leads to 
similar contribution to the cohomology. The total contribution from the 
twisted sector to $H^{\star}(\tilde{X})$ is summarised in the following 
Hodge diamond:
\begin{eqnarray}
\begin{matrix}
&&&&0&&&&\\
&&&0&&0&&&\\
&&0&&6&&0&&\\
&0&&0&&0&&0&\\
0&&0&&13&&0&&0\\
&0&&0&&0&&0&\\
&&0&&6&&0&&\\
&&&0&&0&&&\\
&&&&0&&&&\\
\end{matrix}
\end{eqnarray}
Let us point out that since the contribution from the twisted 
sector is $h^{11}=6$, the number of perturbation parameters 
allowed in the gauge theory in the absence of discrete torsion
is 6. This is in keeping with the fact that there are six deformations 
of the D-flatness condition \cite{ahn}.
Let us point out that there is no symmetry guaranteeing that the contributions 
from the six sectors mentioned above should be the same. That the 
contributions are indeed the same is a coincidence and 
has to be checked on a case by case basis for each of the  elements.
\item {\sl With discrete torsion:}\\
In the presence of discrete torsion, the condition of $G$-invariance of
forms is generalised to \eq{GinvDT}, with $\varepsilon$ defined by 
\eq{epsil}.

The contribution to $H^{\star}(\tilde{X})$ 
corresponding to $(111)$ \emph{happens} to remain the 
same as that in the case without discrete torsion, namely $h^{22}=1$.
 
However, the contribution from the other six do differ.
Considering $(100)$, again, the invariant forms on the 
fixed $\C^2$ are $dZ_3\wedge dZ_4$, $d\bar Z_3\wedge d\bar Z_4$,
$dZ_3\wedge d\bar Z_4$ and 
$d\bar Z_3\wedge dZ_4$. Thus, taking into account the shift by the age 
$s=1$ of each of the elements, 
the contribution to $H^{\star}(\tilde{X})$ is $h^{22}=2$, $h^{31}=1$. 
Analysing similarly the contributions from the other five elements, we 
get, in the long run, the following Hodge diamond arising from the 
twisted sector in the presence of discrete torsion:
\begin{eqnarray}
\begin{matrix}
&&&&0&&&&\\
&&&0&&0&&&\\
&&0&&0&&0&&\\
&0&&0&&0&&0&\\
0&&6&&13&&6&&0\\
&0&&0&&0&&0&\\
&&0&&0&&0&&\\
&&&0&&0&&&\\
&&&&0&&&&\\
\end{matrix}
\end{eqnarray}
\end{itemize}
Thus, we may have at most six deformations for the F-term, as they 
correspond to deformation of the orbifold singularity and determined 
by $h^{31}$. In the next section we will find out the six possible terms.

Let us note that the element $(111)\in G$ has age $s=2$ and leads to a
terminal singularity not giving in to a crepant resolution. However, 
this does not affect the present analysis, as 
this is confined to a consideration of perturbation of the gauge 
theory by the six parameters that correspond to the six (3,1)-forms, 
none of which have been contributed by $(111)$.
Let us point out in passing that by turning on the discrete torsion 
we get the $h^{11}$ and $h^{(4-1)1}=h^{31}$ interchanged. This 
signifies that the resulting manifolds are related  by mirror symmetry.
\section{The gauge theory of the branes}\label{gaugesec}
In the regime of weak string coupling, 
D-branes admit a  description in terms of a supersymmetric 
Yang-Mills theory (SYM) on their world 
volumes. The moduli space of the SYM is interpreted as the 
space-time. Referring to the case at hand, the world-volume
theory of $n$ coalescing 
D1-branes in type-IIB theory 
is taken to be the dimensional reduction 
of the $\mathsf{N} =1$, $\mathsf{D}=10$ $U(n)$ SYM on $\C^4$, 
or equivalently, the reduction of the $\mathsf{N}=4$, $\mathsf{D}=4$ SYM
on $\C^2$,
down to two dimensions. The resulting  theory
is an $\mathsf{N}=(8,8)$, $\mathsf{D}=2$ $U(n)$ SYM.
We will consider a fourfold transverse to the world-volume of 
the D1-branes obtained as an orbifold of $\C^4$.
The D1-brane is taken to be lying  along the 9-th direction, evolving
in time along the 0-th direction. The $\C^4$ is coordinatised by 
$Z_I$, $I=1,2,3,4$.
Let $G$ be a finite group of order $\vert G\vert$. 
In order to retain some supersymmetry, $G$ must be a subgroup of the 
holonomy group of the fourfold, namely $SU(4)$.
The theory of a single D1-brane on the blown down orbifold
$\C^4/G$ is obtained by starting with a theory of 
$\vert G\vert$ coalescing D1-branes in two dimensions
and then quotienting the resulting 
gauge theory with gauge group $U(\vert G\vert)$, by $G$.  
The resulting theory turns out to have $\mathsf{N} =(0,2)$ supersymmetry
in  two dimensions \cite{mohri1}.

It is convenient, in practice, to start with the 
$\mathsf{N} =1$ , $\mathsf{D}=10$ gauge theory
reduced to two dimensions written in a $\mathsf{N}=(0,2)$ 
notation \cite{mohri,wit}. Finally one substitutes the fields those have 
survived the orbifold projection. Through the projection
the supersymmetry will get broken down to $(0,2)$.
This theory corresponds to the theory of a D1-brane on the singular
orbifold $\C^4/G$. 
The deformation and/or blow up of the orbifold $\C^4/G$ 
corresponds to adding extra terms to the above-mentioned (0,2) 
action, with all 
fields taken to be the ones surviving the projection \cite{mohri1}. 
\subsection{The gauge theory: before projection}
Let us begin with an inventory of the multiplets of
$\mathsf{N} = (0,2)$, $\mathsf{D} = 2$  super-Yang-Mills theory
\cite{wit,mohri1,dis}.
The field-content of the theory is as follows. 
There are four complex bosonic fields, denoted by
$Z_I$, $I=1,2,3,4$, identified with the four
complex dimensions transverse to the world 
volume. There are eight left-handed and eight 
right-handed Majorana-Weyl spinors which constitute 
four left-handed Dirac fermions and four right-handed
Dirac fermions. The left handed fermions are $\lambda_-$
satisfying $(\lambda_-)^\dagger = \bar\lambda_-$ 
and the three other can be grouped
together as $\lambda_{IJ}$ which is antisymmetric 
in ${IJ}$ and satisfies 
$(\lambda_{IJ})^\dagger = \epsilon_{IJKL}\lambda_{KL}$.
Finally, there is a vector field, whose two components will
be denoted by $v_\pm$, in the light-cone coordinates.

The fields mentioned above may be assorted into three 
supersymmetry multiplets, that is, into three superfields. 
The vector field and the
fermion $\lambda_-$,  
from the left sector are combined to form the 
{\sl vector multiplet}, whose components in the Wess-Zumino
gauge are written as,
\begin{equation}
\begin{split}
A_- &= v_- - 2i(\theta\bar\lambda_- 
+\bar\theta\lambda_-) + 2(\theta\bar\theta)D, \\
A_+ &= (\theta\bar\theta) v_+,
\end{split}
\end{equation} 
where $v_\pm$ are the vector fields and
$D$ is an auxiliary field.
The corresponding field strength is
given by
\begin{equation}
F = \lambda_- - \theta(F_{-+}+iD) +
2i(\theta\bar\theta)\partial_+\lambda_- ,
\end{equation}
where $F_{-+} = \partial_-v_+ 
- \partial_+v_-$. 

Four {\sl bosonic chiral
multiplets} are formed from $Z_I$ and
four Dirac spinors $\psi_I$ coming from right 
sector.
The corresponding superfield takes the form
\begin{equation}
\phi_I = Z_I + {\sqrt 2}\theta\psi_I
+ 2i\bar\theta\theta \nabla_+ Z_I,
\end{equation}
where we have defined $\nabla_\pm = \partial_\pm - iq A_\pm$,
and $q$ denotes the charge of the vector multiplet under the gauge group
$U(N)$.

Finally, the three fermions $\lambda^{IJ}$,
with hermitian conjugates defined as $(\lambda^{IJ})^\dagger = 
\epsilon^{IJKL}\lambda^{KL}$,
are gathered into three {\sl fermionic chiral multiplets}
which assume the following form:
\begin{equation}
\Lambda_{i4} = \lambda_{i4} -{\sqrt 2}
\theta G_{i4} -2i (\bar\theta\theta)\nabla_+
\lambda_{i4} - {\sqrt 2}\thinspace\bar\theta E_i ;
\quad i=1,2,3,
\end{equation}
where $G_{i4}$ denotes a bosonic auxiliary field 
and $E_i$ represents a bosonic chiral field.

The contributions of these multiplets to the action
of the theory can be obtained \cite{wit} from the 
reduction of ten-dimensional action. A more convenient
way would be to start with four-dimensional 
$\mathsf{N} = 1$ action reduced to two dimensions 
which is a $(2,2)$ theory and then by integrating out
one of the $\theta$'s. These result in fixing the 
chiral fields $E_i$ as $E_i = [\phi_i, \phi_4]$. The 
contributions of the above-mentioned multiplets to the Lagrangian 
of the theory are given by,
\begin{equation}
\begin{split}
L_{A} &= \frac{1}{2e^2} 
\int d\theta d\bar\theta \tr(\bar F F) \\
L_{\phi} &= i \sum_I\int d\theta 
d\bar\theta\tr(\bar\phi_I\nabla_-\phi_I) \\
L_{\Lambda} &= \frac{1}{2}\sum_i
\int d\theta d\bar\theta 
\tr(\bar\Lambda_{i4}\Lambda_{i4i})
\end{split}
\label{3action}
\end{equation}

The total Lagrangian  obtained as the sum of the three pieces \eq{3action}
admits a superpotential term, while retaining $\mathsf{N}=(0,2)$
supersymmetry. The corresponding piece of the Lagrangian is given by  
\begin{equation}
L_{\mathcal W}= \frac{1}{\sqrt 2} \int d\theta~{\mathcal W} + 
\text{h.c.,}
\end{equation}
where $\mathcal W$ is a chiral fermionic field, the 
\emph{superpotential}. The general form of $\mathcal W$ is 
$\mathcal W = \tr\sum_i\Lambda_{i4}J^i$,
where $J^i$ denotes a bosonic chiral field satisfying
the supersymmetry-constraint
\begin{equation}
\sum_iE_iJ^i = 0. \label{dsusy}
\end{equation}
Thus, to sum up, the total action 
is given by 
\begin{equation}\label{totlag}
L = L_{A} + L_{\phi} + L_{\Lambda} + L_{\mathcal W}.
\end{equation}
The reduced two-dimensional theory, 
in absence of extra couplings, the chiral field $J^i$ in $\mathcal W$
assumes the form, $J^i = \sum_{j,k}\epsilon^{ijk}[\phi_j, \phi_k]$, which
satisfies the supersymmetry constraint \eq{dsusy}, thanks to 
the Jacobi identity.

The action corresponding to \eq{totlag} has a global $U(1)^4$ symmetry
associated with the phases of the bosonic fields of which the global
$U(1)$ is an R-symmetry of the theory. 
The bosonic potentials, known as the F-term and the D-term,
are obtained 
by integrating out the auxiliary fields $G_{i4}$ and $D$, respectively,
appearing in $L$. These are given by 
\begin{eqnarray}
U_F &=& 2 \sum_{I,J} \tr[Z_I, Z_J]^2, \\\label{d-flatness} 
U_D &=& 2e^2 \sum_I \tr[Z_I , \bar Z_I], 
\end{eqnarray}
respectively.

As we will see in presence of the coupling of D-string 
to twisted sector modes the superpotential as well as the
form the chiral fields $E_i$ will get modified.
\subsection{The gauge theory: after projection}
Having discussed some generalities 
let us now implement the projection by  
$G = \Z_2\times\Z_2\times\Z_2$. The order of $G$ is $\vert G\vert = 8$. 
Hence we start with 8 branes at the origin, choosing the gauge 
group to be $U(8)$, and then quotient the theory by $G$.
As mentioned earlier, the action of the group on the Chan-Paton indices is 
given by the regular representation, obtained as direct sum of 
two copies of each of the projective representations given in 
\eq{REP}. Thus, we have, 
\begin{equation}
r(g_i) = \diag{\sigma_i, \sigma_i, -\sigma_i, -\sigma_i},
\end{equation}
where $r(g_i)$ denotes the regular representation of the 
generator $g_i$ of the group $G$.
The action of the generators $g_i$ on the Lorentz indices is as given in 
\eq{sptimeact}. 
              
The projections are given by adjoint action of the regular representations
$r(g_i)$ on the supermultiplets. In terms of 
the respective superfields, they take the form, 
\begin{equation} \label{proj1}
\begin{split}
r(g_i)A_\mu r(g_i)^{-1} &= A_{\mu}, \\
r(g_i)Z_I r(g_i)^{-1} &= \sum_{I=1}^4 R(g_i)_{IJ}Z_J, \\
r(g_i)\lambda_{IJ} r(g_i)^{-1} &= \sum_{I',J' =1}^4 
R(g_i)_{II'}R(g_i)_{JJ'}\lambda_{I'J'},
\end{split}
\end{equation}
where we have introduced three matrices $R(g_i)_{IJ}$, such that the equations
\eq{sptimeact} read 
\begin{equation}
g_i:~Z_I \map \sum_{J=1}^4R(g_i)_{IJ}Z_J, 
\end{equation}
for $I=1,2,3,4$ and $i=1,2,3$.

The gauge group of the theory after projections \eq{proj1} breaks down to
$U(2)\times U(2)$ 
of which the center of mass $U(1)$ decouples which
plays the role of the unbroken $U(1)$ for the single 
brane. The rest of the group $U(1)\times SU(2)\times 
SU(2)$ is broken by the vacuum expectation 
values of the following Higgs field
\begin{equation}\label{higgs}
Z_I = \left(\begin{array}{cc} 0 & z_I\times\sigma_I\\
w_I\times\sigma_I & 0 \end{array}\right),
\end{equation}
where $z_I$, $w_I$ are $2\times 2$ matrices.
The charges  are assigned through,
\begin{equation} \label{trans}
z_I \map U z_I V^{\dagger}, \qquad 
w_I \map V w_I U^{\dagger}
\end{equation}
where $U$ and $V$ belong to the two $SU(2)$'s
and $z_I$ and $w_I$ have opposite charges under
the relative $U(1)$.
The representation of the chiral Fermi field is the same
as that of the commutators of the bosonic fields, which
is consistent with the presence of the $E$ field in the 
multiplet. Thus, we have
\begin{equation}
\Lambda_{i4} = \left(\begin{matrix} \lambda^1_i\times\sigma_i & 0 \\ 
0 & \lambda^2_i\times\sigma_i \end{matrix} \right).
\end{equation}

The fields after projection are then substituted in the 
action in order to 
derive the theory of the single brane on $\C^4/G$. 
The moduli space is described by the solutions of the conditions 
of F-flatness and D-flatness, obtained by 
minimising the F-term and the D-term, respectively.
The F-flatness conditions are given by
\begin{equation}\label{ud}
\begin{split}
z_j w_i + z_i w_j &= 0,\qquad 
z_4 w_i - z_i w_4 = 0, \\
w_j z_i + w_i z_j &= 0,\qquad
w_4 z_i - w_i z_4 = 0, 
\end{split}
\end{equation}
for $i=1,2,3$.

Let us now consider the deformation of 
this moduli space in the presence of 
non-zero coupling with fields in the 
twisted sector of the closed string theory. In the 
discussion of the closed string twisted 
sectors in \S\ref{closed}, we noticed that each of the six 
group elements which flips the sign of 
$Z_I$'s pairwise contributes $(3,1)$, 
responsible for the deformation of 
the complex structure 
of the singular orbifold. These are complex 
numbers and hence 
couple naturally to the superpotential
\cite{dt1,dt2}. Geometrically, these deform 
away codimension-two singularities.

One set of the natural modification of the
superpotential comes from its four-dimensional analogue.
Starting with a term of the form $\int d\theta \xi_i\phi_i$, 
where $\phi_i$ is a four-dimensional chiral field, reducing
to two dimension and integrating one $\theta$ out
yields a term
\begin{equation}
\delta W = \sum_{i=1}^3\int d\theta^+\tr(\boldsymbol{\xi}_i\Lambda_{i4}),
\end{equation}
where $\boldsymbol{\xi}_i$ denotes the coupling parameters, of the form
\begin{eqnarray}
\boldsymbol{\xi}_i = \begin{pmatrix} \xi_i\times
\sigma_i & 0 \\ 
0 & \xi_i\times\sigma_i 
\end{pmatrix}.
\end{eqnarray}
The form of $\boldsymbol{\xi}_i$ is determined
by the gauge-invariance of the coupling and by the
fact that its introduction does not break
supersymmetry according to (\ref{dsusy}). A simple calculation leads to the
above expression without any loss of generality,
provided $\boldsymbol{\xi}_i$ do not depend
on the other fields.

These perturbations give rise to a deformation 
of the F-term equation as
\begin{equation}
\begin{split}
\label{d1} 
z_j w_i + z_i w_j &= e_{ijk}\xi_k,  \\
w_j z_i + w_i z_j &= e_{ijk}\xi_k,    
\end{split}
\end{equation}
where $i=1,2,3$, and $e_{123}=+1$, $e_{ijk}$ is 
symmetric under interchange of $i,j$ but non zero 
only when $i,j,k$ are all different.

The form of the other perturbations can be
obtained by treating all the 4 coordinates
of the transverse space on same footing.
This perturbations can be introduced by making
use of the freedom in the definition of $E$
which correspond to a perturbation
of the field $E_i$ in the fermionic multiplet
given by
\begin{equation}\label{pert}
E_i \rightarrow E_i + \boldsymbol{\eta}_i,
\end{equation}
where 
\begin{eqnarray}
\boldsymbol{\eta}_i = \begin{pmatrix} \eta_i\times
\sigma_i & 0 \\ 
0 & \eta_i\times\sigma_i 
\end{pmatrix} .
\end{eqnarray}
Here also the form of the coupling is determined by 
the fact that $\boldsymbol{\eta}_i$ should transform in a similar
way as $E_i$ and it should not break supersymmetry
(\ref{dsusy}).

To sum up, the F-term equations following from these
perturbations are given by
\begin{equation}
\label{f-flatness}
\begin{split}
z_j w_i + z_i w_j &= e_{ijk}\xi_k, \qquad
z_4 w_i - z_i w_4 = \eta_i, \\
w_j z_i + w_i z_j &= e_{ijk}\xi_i,\qquad
w_i z_4 - w_4 z_i  = \eta_i.
\end{split}
\end{equation}
We shall sometimes refer to \eq{f-flatness}
as the \emph{F-flatness equations}.
Let us note that, we have six parameters 
$\xi_i$, $\eta_i$, $i=1,2,3$,
appearing in the perturbation of the gauge 
theory. This is in keeping 
with the fact that, the contribution to 
the cohomology of the 
resolved space, arising in the twisted 
sector is $h^{31}=6$, as
discussed in \S\ref{closed}. 
\section{The moduli space and its deformation}\label{secdeform}
Let us now go over to finding the vacuum moduli space of the 
(0,2) theory discussed in the previous section. The vacuum moduli 
space is the space of allowed values of the scalars $Z_I$, respecting 
the F- and D-flatness conditions, up to gauge equivalence.
\subsection{Gauge-invariant polynomials}
Thus, we have the four matrices $Z_I$, in the form \eq{higgs}, and 
the F-flatness conditions \eq{f-flatness}
on the non-vanishing $2\times 2$ blocks of $Z_I$, namely $z_I$ and 
$w_I$. The equation of the variety describing the moduli space of 
the configuration will be written in terms of the gauge-invariant 
polynomials constructed out of $z_I$ and $w_I$. We proceed to 
describe these next. 

Let  us introduce the following expressions,
\begin{equation}
\label{defnP}
\begin{split}
\mathcal P_{IJK\cdots} &= \frac{1}{2}\tr{} z_Iw_Jz_K\cdots ,\\
\tilde{\mathcal P}_{IJK\cdots} &= \frac{1}{2}\tr{} w_Iz_Jw_K\cdots ,
\end{split}
\end{equation}
which we will refer to as \emph{polynomials}. By the 
\emph{order} of a polynomial, we mean the total number of $z$ and $w$ 
appearing in the polynomial --- that is, the length 
of the word inside the trace. We need to introduce a few further notations
which we list here:
\begin{eqnarray}
\label{QQ}
\mathcal Q_{IJ} = z_Iw_J,\qquad 
\tilde{\mathcal Q}_{IJ} = w_Iz_J,\\\label{xx}
x_I = \frac{1}{2}\tr{\mathcal Q_{II}},\qquad 
\tilde{x}_I = \frac{1}{2}\tr{\tilde{\mathcal Q}_{II}}. 
\end{eqnarray}
Let us also note that, 
\begin{equation}
\tilde{x}_I = x_I.
\end{equation}
It then follows from \eq{defnP} that
\begin{equation}
\mathcal P_{IJ} = \frac{1}{2}\tr{\mathcal Q_{IJ}} \qquad 
\tilde{\mathcal P}_{IJ} = \frac{1}{2}\tr{\tilde{\mathcal Q}_{IJ}} 
\end{equation}
Now, the polynomials of different orders defined by \eq{defnP} are not
linearly independent. 
We need to find out the independent polynomials in order to write down 
the equation of the variety.
At this point let us note that from the gauge transformation \eq{trans} 
it follows that only the polynomials of even order are gauge-invariant 
quantities. Hence in what follows we shall not consider 
the polynomials of odd orders. 
Let us consider the remaining polynomials order by order.
\begin{itemize}
\item \underline{Order 2 polynomials}\\
Using the constraints \eq{f-flatness}, it can be shown that $\mathcal Q_{II}$
commute pairwise. Hence, assuming that these matrices are non-singular, 
the matrices $\mathcal Q_{II}$ can be simultaneously diagonalised. From 
now on we assume that $\mathcal Q_{II}$, $I=1,2,3,4$  are diagonal.
Moreover, $\mathcal Q_{IJ}$ with $I\neq J$ commute
with $\mathcal Q_{II}$, but not between themselves; for example, 
$\mathcal Q_{12}$ and $\mathcal Q_{23}$ do not commute. Hence, in a 
basis in which $\mathcal Q_{II}$ are diagonal, $\mathcal Q_{IJ}$, with 
$I\neq J$ are necessarily generic, \ie not diagonal. 
It then follows, from the fact that $\mathcal Q_{IJ}$, with $I\neq J$ 
and $\mathcal Q_{II}$ commute, that $\mathcal Q_{II}$ are all 
proportional to the two-dimensional identity matrix, $\id{2}$. 
Moreover, if $z_I$ and $w_I$ are non-singular matrices, which also 
we assume, it follows that $\tilde{\mathcal Q}_{II}$ is also 
proportional to $\id{2}$. Thus we have, 
\begin{equation}\label{diagonal}
\mathcal Q_{II}=\tilde{\mathcal Q}_{II} = x_I\id{2},
\end{equation}
and consequently,
\begin{equation}\label{diagp}
\mathcal P_{II} = x_I.
\end{equation}
The F-flatness conditions \eq{f-flatness} impose further
constraints on $\mathcal Q_{IJ}$ with $I\neq J$. We will use these later.

Since the polynomials ${\mathcal P}_{IJ}$ and $\tilde{\mathcal P}_{IJ}$
satisfy $\tilde{\mathcal P}_{IJ}= {\mathcal P}_{JI}$,  it 
suffices to consider either set.
The twelve polynomials $\mathcal P_{IJ}$ with $I\neq J$ are
not linearly independent. 
The F-flatness conditions \eq{f-flatness} reduce these to a set of 
three independent ones. There are three further relations among 
these six, namely,
\begin{equation}\label{constraint}
\begin{split}
\xi_2\mathcal P_{24} - \xi_3\mathcal P_{34} + \eta_2\mathcal P_{13} -
\eta_3\mathcal P_{12} &= 0,\\
\xi_1\mathcal P_{14} - \xi_2\mathcal P_{24} - \eta_1\mathcal P_{23} + 
\eta_2\mathcal P_{13} &= \xi_2\eta_2 -\xi_1\eta_1,\\ 
\xi_1\mathcal P_{14} - \xi_3\mathcal P_{34} + \eta_1\mathcal P_{23} +
\eta_3\mathcal P_{12} &= \xi_3\eta_3.
\end{split}
\end{equation}
The equations \eq{constraint} can be derived by considering certain 
polynomials of order 4 and using properties of trace of products of 
matrices and \eq{f-flatness}. For example, the first equation in 
\eq{constraint} follows from
$\mathcal P_{1243}$ as,
\begin{equation}
\begin{split}
\mathcal P_{1243} &= \frac{1}{2}\tr{\zpol{1234}{4}}\\
 &= \frac{1}{2}\tr{(\xi_3 - z_2w_1)(z_3w_4 + \eta_3)}\\
 &= \xi_3\mathcal P_{34} - \eta_3\mathcal P_{21} + \xi_3\eta_3 - 
\tilde{\mathcal P}_{1342}\\
\tilde{\mathcal P}_{1342} &= \frac{1}{2}\tr{\wpol{1342}{4}}\\
&= \frac{1}{2}\tr{(\xi_2 - w_3z_1) (-\eta_2+w_2z_4)}\\
&= -\xi_2\eta_2 + \eta_2\mathcal P_{13} + \xi_2\mathcal P_{42} - 
\mathcal P_{1243},
\end{split}
\end{equation}
and further using \eq{f-flatness}. 
The second and third relations follow similarly by considering 
$\mathcal P_{1243}$ (again!) and $\mathcal P_{2341}$ respectively.
Thus, finally we are left with seven gauge-invariant polynomials 
of order two.
\item\underline{Order 4 polynomials}\\
Three cases arise for the polynomials $\mathcal P_{IJKL}$.
\begin{enumerate}
\item $\mathcal P_{IIJK}$ for all $I$, $J$, $K$.  These are determined 
in terms of $\mathcal P_{II}$ and $\mathcal P_{JK}$ by virtue of 
\eq{diagonal}. 
\item $\mathcal P_{IIJJ}= \frac{1}{2}x_Ix_J$
\item $\mathcal P_{1234}$ --- this is an independent polynomial. Other 
order 4 polynomials with all indices different are determined in terms 
of $\mathcal P_{1234}$ and certain polynomials of order 2.
\end{enumerate}
As for the polynomials $\tilde{\mathcal P}_{IJKL}$, 
\begin{enumerate}
\item $\tilde{\mathcal P}_{IIJK}$ for all $I$, $J$.  These are determined 
in terms of $\tilde{\mathcal P}_{II}$ and $\tilde{\mathcal P}_{JK}$ 
by virtue of \eq{diagonal}, and the latter are related to  
${\mathcal P}_{II}$ and ${\mathcal P}_{JK}$ in turn, again by \eq{diagonal}. 
\item $\tilde{\mathcal P}_{IIJJ}$ are also determined in terms of the 
are determined by $x_I$
\begin{equation}\begin{split}
\tilde{\mathcal P}_{IIJJ}&= \frac{1}{2}\tilde{x}_I\tilde{x}_J \\
&= \frac{1}{2}x_Ix_J.\notag
\end{split}\end{equation}
\item $\tilde{\mathcal P}_{1234}$ is determined in terms of 
$\mathcal P_{1234}$ by the relation 
\begin{equation}
\tilde{\mathcal P}_{1234} =  \mathcal P_{1234} 
+ \frac{1}{2}(\xi_1\eta_1 - \xi_2\eta_2+\xi_3\eta_3),
\end{equation}
as a consequence of the F-flatness conditions \eq{f-flatness}
\end{enumerate}
Thus,  we conclude that, the only independent polynomial of 
order four is $\mathcal P_{1234}$.
\item\underline{Polynomials of order higher than 4}\\
These can all be expressed in terms of polynomials of lower order.
Some of these relations follow by \eq{diagonal} and \eq{diagp}, while 
the others follow by an use of the Cayley--Hamilton theorem. 
As an example of the latter cases, let us consider the polynomial 
$\mathcal P_{111134}$
= $\frac{1}{2}\tr{\zpol{111134}{6}}$. The Cayley-Hamilton theorem 
leads to the following equation for any $2\times 2$ matrix, $\mathbf M$.
\begin{equation}
\mathbf{M}^2 - \mathbf{M}\tr{\mathbf M} + \frac{1}{2}\Big((\tr{\mathbf M})^2 - 
\tr{\mathbf M}^2 \Big)\id{2} = 0.
\end{equation}
Now, taking $\mathbf M = z_1w_1$, we have 
\begin{equation}
\zpol{1111}{4} - 2\zpol{11}{2}\mathcal P_{11} 
+ \Big(2\mathcal P_{11}^2 - \mathcal P_{1111}\Big)\id{2} = 0.
\end{equation}
Multiplying by $\zpol{34}{2}$ and taking trace we obtain
\begin{equation}
\mathcal P_{112234} = 2\mathcal P_{11}\mathcal P_{1134} -
2\mathcal P_{11}^2\mathcal P_{34} + 
\mathcal P_{1111}\mathcal P_{34}.
\end{equation}
Thus, the polynomial $\mathcal P_{112234}$ of order six 
is expressed in terms of polynomials of lower order.
\end{itemize}
Other polynomials of order higher than four can be treated 
similarly. In particular,
in any polynomial  of order say $(4+k)$, $k>0$,
at least $k$ indices come repeated.
Using (\ref{f-flatness}) it is always possible to gather these repeated 
indices together and use the proportionality of 
$\mathcal Q_{II}$  to the identity matrix to reduce it to 
some combination of lower order polynomials or use the Cayley-Hamilton 
theorem as above.
Thus, finally we have ten independent, gauge invariant polynomials 
of order two and one of order four. The vacuum moduli space is described 
by an equation involving these polynomials upon using the 
F-flatness conditions.
\subsection{The variety}
Now, in order to write down the equation for the variety describing 
the moduli space, let us introduce a $4\times 4$ matrix $\mathcal M$, 
defined as follows.
\begin{equation}
\begin{split}
\mathcal M_{II} &= \mathcal P_{II},\\
\mathcal M_{ij} &= \frac{1}{2}\Big(\mathcal P_{ij}+
\tilde{\mathcal P}_{ij}\Big),\\
\mathcal M_{i4} &= \frac{1}{2}\Big(\tilde{\mathcal P}_{i4} - 
\mathcal P_{i4}\Big),\\
\mathcal M_{4i} &= \frac{1}{2}\Big(\tilde{\mathcal P}_{4i}-
{\mathcal P}_{4i}\Big)\\ &= -\mathcal M_{i4},
\end{split}
\end{equation} 
for $I,J = 1,2,3,4$ ans $i,j = 1,2,3$.
Defining another complex variable $t$ as 
\begin{equation}
t =\frac{1}{2i} \Big(\mathcal P_{1234} + \tilde{\mathcal{P}}_{1234}\Big), 
\end{equation}
the equation of the variety is written as 
\begin{equation}\label{dett}
t^2 = \det \mathcal M. 
\end{equation}
We now write this in terms of $t$ and $x_I$, $I=1,2,3,4$ and use
\eq{f-flatness}. The matrix $\mathcal M$ now 
assumes the following form:
\begin{eqnarray}\label{defnM}
\mathcal M =\begin{pmatrix}
x_1 & \xi_3/2 & \xi_2/2 &\eta _1/2\\
\xi_3/2&x_2 & \xi_1/2 &\eta_2/2\\
\xi_2/2&\xi_1/2&x_3&\eta_3/2\\
-\eta_1/2&-\eta_2/2&-\eta_3/2&x_4\\
\end{pmatrix}.
\end{eqnarray}
The matrix $\mathcal M$ is the most general one consistent
with the global $U(1)^4$ symmetry, provided we 
assign the correct charges to the parameters $\xi$ 
and $\eta$ from the action. 

The relation \eq{dett} can be argued from the
algebra of the matrices $\mathcal Q_{IJ}$ that
follows from \eq{f-flatness}. Any solution
is a represenattion of the algebra and it
can be shown from the algebra that 
in the generic case the
moduli are related through \eq{dett}.

Hence the equation of the variety becomes 
\begin{equation}\label{variety}
\mathcal F(x_1,x_2,x_3,x_4,t) = 0,
\end{equation}
where we have defined
\begin{equation}\label{defF}
\begin{split}
\mathcal F(x_1,x_2,x_3,x_4,t) =  x_1x_2x_3x_4 
&- \frac{1}{16}(\xi_1^2\eta_1^2 + \xi_2^2\eta_2^2 + \xi_3^3\eta_3^2)\\
&+ \frac{1}{8}(\xi_1\xi_2\eta_1\eta_2 + \xi_2\xi_3\eta_2\eta_3 +
\xi_3\xi_1\eta_3\eta_1) \\
&- \frac{1}{4}(\xi_1\eta_2\eta_3x_1 + \xi_2\eta_3\eta_1x_2  
      +\xi_3\eta_1\eta_2x_3) \\
&+ \frac{1}{4}(\eta_3^2x_1x_2 + \eta_1^2x_2x_3 + \eta_2^2x_3x_1)\\
&- \frac{1}{4}(\xi_1^2 x_1 + \xi_2^2 x_2 + \xi_3^2 x_3 - \xi_1\xi_2\xi_3)x_4\\ 
&-t^2.
\end{split}
\end{equation}
That the equation \eq{variety} is indeed satisfied by solutions of 
the F-flatness conditions should be explicitly checked 
on any solution of \eq{f-flatness}. 
Let us verify this in some examples.
\renewcommand\labelitemi{\null}
\begin{itemize}
\item{\bf Example~1}\\
When $\xi_i=\eta_i=0$ for $i=1,2,3$, a 
solution of \eq{f-flatness} is given by 
\begin{equation}\label{soln1}
\begin{split}      
z_I  &= \mathsf{z}_I\sigma_I\\
w_I  &= \mathsf{w}_I\sigma_I,\quad I=1,2,3,4,
\end{split}
\end{equation}
where $\mathsf{z}_I $ and $\mathsf{w}_I$ are complex numbers.
The F-flatness constraints \eq{f-flatness} impose the following 
relations among $\z_I$ and $\w_I$:
\begin{equation}
\begin{split}\label{fflat}
\z_1\w_2 = \z_2\w_1,  &\quad \z_1\w_4 = \z_4\w_1, \\
\z_2\w_3 = \z_3\w_2,  &\quad \z_2\w_4 = \z_4\w_2, \\
\z_3\w_1 = \z_1\w_3,  &\quad \z_3\w_4 = \z_4\w_3.
\end{split}
\end{equation}

We also need the constraints ensuing from the D-flatness
condition \eq{d-flatness}, namely
\begin{equation}\label{dflatt}
|\z_1|^2+|\z_2|^2+ |\z_3|^2+|\z_4|^2
-|\w_1|^2-|\w_2|^2-|\w_3|^2-|\w_4|^2=0.
\end{equation}
Let us note that only the three relations  written in the  second column
of \eq{fflat} are independent --- the other three relations in the first
column follow from these. 
Also, note that the three equations in the second column of \eq{fflat}
are monomial equations. Therefore, we can give a toric description 
to the envisaged variety, following \cite{dgm}. In the notation of 
\cite{subir},
we can express the variables $\z_I$, $\w_I$, in terms of the five 
independent variables, which we choose to be $\z_1$, $\z_2$, $\z_3$, $\z_4$
and $\w_4$. This is expressed as:
\begin{eqnarray}
\bordermatrix{
&\z_1 & \z_2 &\z_3 &\z_4&\w_1&\w_2&\w_3&\w_4\nonumber\\
\z_1& 1&0&0&0&1&0&0&0\nonumber\\
\z_2& 0&1&0&0&0&1&0&0\nonumber\\
\z_3& 0&0&1&0&0&0&1&0\nonumber\\
\z_4& 0&0&0&1&-1&-1&-1&0\nonumber\\
\w_4& 0&0&0&0&1&1&1&1\nonumber
}.
\end{eqnarray}
In terms of six homogeneous coordinates, $p_i$, $i=0,\cdots 5$,
the five independent variables can then be expressed as:
\begin{eqnarray}
T=\bordermatrix{
&p_0&p_1&p_2&p_3&p_4&p_5\nonumber\\
\z_1& 1&1&0&0&0&0\nonumber\\
\z_2& 1&0&1&0&0&0\nonumber\\
\z_3& 1&0&0&1&0&0\nonumber\\
\z_4& 1&0&0&0&1&0\nonumber\\
\w_4& 0&0&0&0&1&1\nonumber
}.
\end{eqnarray}
The kernel of $T$ is given by  
\begin{eqnarray}\label{kert}
(\ker{T})^{\mathrm{T}} = \begin{pmatrix}1&-1&-1&-1&-1&1\end{pmatrix},
\end{eqnarray}
and provides part of the charge matrix for the variety.
We can also find out the charges of the five independent variables 
under a $\C^{\star}$ from the D-flatness condition \eq{dflatt} as
\begin{eqnarray}
V= \begin{pmatrix} 1 & 1& 1& 1& -1\end{pmatrix}.
\end{eqnarray}
The charges of the homogeneous coordinates $p_i$ under this 
$\C^{\star}$  are obtained 
from $V$, using a matrix $U$, satisfying $TU^{\mathrm{T}} = \id{5}$, namely,
\begin{eqnarray}
U=\begin{pmatrix}
0&1&0&0&0&0\\
0&0&1&0&0&0\\
0&0&0&1&0&0\\
1&-1&-1&-1&0&0&\\
0&0&0&0&0&1
\end{pmatrix},
\end{eqnarray}
as follows: 
\begin{eqnarray}
VU = \begin{pmatrix}
1&0&0&0&0&-1
\end{pmatrix}.
\end{eqnarray}
Concatenating this with $(\ker{T})^{\mathrm T}$ in \eq{kert}, we obtain
the charge matrix
\begin{eqnarray}
\tilde{Q} = \begin{pmatrix}
1&-1&-1&-1&-1&1\\
1&0&0&0&0&-1
\end{pmatrix}.
\end{eqnarray}
The co-kernel of the transpose of $\tilde{Q}$, after deleting
a column, identical to the first one, takes the form
\begin{eqnarray}
\tilde{T} = \coker{\tilde{Q}^{\mathrm{T}}} = \begin{pmatrix}
1&2&0&0&0\\
1&0&2&0&0\\
1&0&0&2&0\\
1&0&0&0&2
\end{pmatrix},
\end{eqnarray}
and is the toric data of the variety under consideration.
The equation of the variety can be read of from the 
kernel of $\tilde{T}$, namely
\begin{eqnarray}
\ker{\tilde{T}}=\begin{pmatrix}
2\\-1\\-1\\-1\\-1
\end{pmatrix}.
\end{eqnarray}
Correspondingly, the equation of the variety under consideration 
is given by 
\begin{equation}
t^2 = x_1x_2x_3x_4. 
\end{equation}
which is the same as \eq{variety} with $\xi_i=\eta_i=0$.  
The five variables appearing in the above equation 
are expressed in terms of the gauge-invariant combinations 
described earlier, as follows.
\begin{equation}\label{soln1xt}
\begin{split}
x_I &= \mathsf{z}_I\mathsf{w}_I,\quad I=1,2,3,4, \\
t &= \frac{1}{2} ({\mathsf z}_1{\mathsf w}_2{\mathsf z}_3{\mathsf w}_4 +
{\mathsf w}_1{\mathsf z}_2{\mathsf w}_3{\mathsf z}_4 ).
\end{split}
\end{equation}
Thus, we have verified the claim that \eq{variety} is indeed the equation of
the variety describing the vacuum moduli space. 
\hfill\ding{113}
\end{itemize} 

\noindent Let us now verify the claim in the case where all the 
deformation parameters are non-vanishing.
We wish to have a solution which goes over to the solution given in
the previous example in the limit of vanishing deformation parameters. 

\begin{itemize}
\item{\bf Example~2}\\
Let us present the solution first.
\begin{equation}
\begin{split}\label{soln2}
z_1 = \mathsf{z}_1\sigma_1 + \xi_3\sigma_2/2\mathsf{w_2},&\quad
w_1 = \mathsf{w}_1\sigma_1 + \xi_3\sigma_2/2\mathsf{z_2},\\
z_2 = \mathsf{z}_2\sigma_2 + \xi_1\sigma_3/2\mathsf{w_3},&\quad
w_2 = \mathsf{w}_2\sigma_2 + \xi_1\sigma_3/2\mathsf{z_3},\\
z_3 = \mathsf{z}_3\sigma_3 + \xi_2\sigma_1/2\mathsf{w_1},&\quad
w_3 = \mathsf{w}_3\sigma_3 + \xi_2\sigma_1/2\mathsf{z_1},\\
z_4 = \z_4\sigma_4+\alpha_1\sigma_1  + \alpha_2\sigma_2+
\alpha_3\sigma_3, &\quad
w_4 = \w_4\sigma_4+\ell(\alpha_1\sigma_1  + 
\alpha_2\sigma_2+\alpha_3\sigma_3), 
\end{split}
\end{equation}
where $\mathsf{z}_I$ and $\mathsf{w}_I$ satisfy the equations 
\eq{fflat} and 
the three complex  variables $\alpha_1$, $\alpha_2$, $\alpha_3$ 
solve the following set of three relations:
\begin{equation}
\begin{split}
2\z_1\w_3\alpha_3 + \xi_2\alpha_1 - \eta_3\z_1 &=0,\\
2\z_2\w_1\alpha_1 + \xi_3\alpha_2 - \eta_1\z_2 &=0,\\
2\z_3\w_2\alpha_2 + \xi_1\alpha_3 - \eta_2\z_3 &=0, 
\end{split}
\end{equation}
and we have defined
\begin{equation}
\ell = -\frac{\w_1}{\z_1} = -\frac{\w_2}{\z_2} = -\frac{\w_3}{\z_3}
= -\frac{\w_4}{\z_4},
\end{equation}
by \eq{fflat}.
It should be noted that $z_I$ and $w_I$ as
given in \eq{soln2}, satisfy the F-flatness conditions \eq{f-flatness}
if $\z_I$ and $\w_I$ satisfy \eq{fflat}. 
Hence, as above, the corresponding variety is again a fourfold,
given by \eq{variety} in $\C^5$. However, the D-flatness conditions alter 
from \eq{d-flatness}.
In terms of the above solution, we have
\begin{equation}\label{t1234}
\begin{split}
x_1 &= \z_1\w_1 + \frac{\xi_3^2}{4\z_2\w_2},\\ 
x_2 &= \z_2\w_2 + \frac{\xi_1^2}{4\z_3\w_3},\\ 
x_3 &= \z_3\w_3 + \frac{\xi_2^2}{4\z_1\w_1},\\ 
x_4 &= \z_4\w_4 + \ell(\alpha_1^2+\alpha_2^2+\alpha_3^2),\\
t   &=  \left(\z_1\w_2\z_3\w_4 + 
\frac{\xi_1\xi_2\xi_3\w_4}{8\w_1\z_2\w_3} \right).
\end{split}
\end{equation}
We have used \eq{fflat} in writing the expression for $t$ in \eq{t1234}.
The equation \eq{variety} is satisfied 
for these values of the five variables.

Let us note that, as mentioned above, we have the same equations 
among the different variables $\z_I $ and $\w_I$ in the present case,
as in Example~1 above. However, the D-flatness conditions are now 
different and involves the parameters $\xi_i$ and $\eta_i$. We have not
solved the D-flatness equations. However, the above solution should be 
gauge-equivalent to a solution that simultaneously solves the 
F- and D-flatness equations \cite{luty}. The variety, anyway, is 
different from the one in the previous example. 
\hfill\ding{113}
\end{itemize}
\subsection{Special branches}
Returning  to the equation \eq{variety} for the variety,
it has a variety of singularities depending on the values 
of the deformation parameters $\xi_i$ and $\eta_i$. 
The singular subsets of the variety \eq{variety} are 
simultaneous solutions of $\mathcal F =0$ and $\pa\mathcal F =0$, 
as noted in \S\ref{closed}. We list some of the cases below.

\noindent{\bf NB:}~~ In this subsection we have rescaled the parameters 
$\xi_i$ and $\eta_i$ to $2\xi_i$ and $2\eta_i$ respectively, in order 
to avoid clumsy factors in the expressions.  
 
\begin{enumerate}
\item In the limit of vanishing deformation, $\xi_i=\eta_i=0$, we recover 
the singular orbifold. This has a $\Z_2\times\Z_2\times\Z_2$ singularity 
at the origin, along with higher dimensional singular subspaces, all 
of which contain the origin.
\item When all $\xi_i =0$, we can have one, two or all 
three of the $\eta_i$ non-vanishing. 
\begin{enumerate}
\item $\eta_i\neq 0$. This has a line singularity
along $x_4$, with $x_1=x_2=x_3=t=0$.
\item $\eta_1=0$, $\eta_2, \eta_3\neq 0$. This also has 
a line singularity along $x_4$, with $x_i= t=0$.
\item $\eta_1$, $\eta_2\neq 0$, $\eta_3=0$. This has a singular plane
given by $x_3=0$,   
$ x_1x_2x_4 + \eta_1^2x_2 + \eta_2^2x_1 = 0$.
\item $\eta_1=\eta_2=0$, $\eta_3\neq 0$. The singularity is
along the union of the two planes given by: $x_3$-$x_4$ plane, 
with $x_1=x_2=t=0$, and the plane, $x_1=0$, $x_3x_4+\eta_3^2=0$, $x_2$
arbitrary.
\end{enumerate}
\item The next cases are when one of the $\xi_i$ is non zero. Let us 
assume, $\xi_1=\xi_2=0$, $\xi_3\neq 0$.  we can have different 
numbers and combinations of $\eta_i$ non-vanishing.
\begin{enumerate}
\item $\eta_i\neq 0$. This has a line
singularity given by 
\begin{equation}\label{spbr1}
x_1 = \frac{\xi_3\eta_1}{\eta_2},\quad
x_2 = \frac{\xi_3\eta_2}{\eta_1}, \quad
x_3x_4 + \frac{\eta_1\eta_2}{\xi_3}x_3 + \eta_3^2 =0
\end{equation}
\item $\eta_1,\eta_2\neq 0$, $\eta_3 = 0$. 
This has a singular line which can be obtained by setting 
$\eta_3=0$ in \eq{spbr1}.
The case with $\xi_i\ne0$, $\eta_i=0$ will be similar.
\item \label{eta13} $\eta_1= 0$ and  $\eta_2,\eta_3\neq 0$.
There is no singularity. This can also 
be seen by taking the limit $\eta_1 \rt 0$ in \eq{spbr1}, 
--- the singularity is sent to infinity along $x_2$.
\item The case with $\eta_2=0$, $\eta_1,\eta_3\neq 0$ is similar to 
(\ref{eta13}).
\item $\eta_1\neq 0$, $\eta_2 =\eta_3 =0$. This has 
a line singularity along $x_1$ with $x_2=x_3=x_4=0$.
This is similar to the case with 
$\eta_2\ne 0$, $\eta_3\ne 0$ and the rest of the parameters vanishing.
\item $\eta_1=\eta_2=0$, $\eta_3\neq 0$. This has singularity
along a plane given by the common solution of
\begin{equation}
x_1x_2-\xi_3^2=0,\qquad x_3x_4+\eta_3^2=0.
\end{equation}
\end{enumerate}
\item Next, let us consider the cases, when two of the $\xi_i$ are
non-vanishing, say $\xi_1$, $\xi_2\neq 0$, $\xi_3 = 0$.
\begin{enumerate}
\item When all $\eta_i\neq 0$, the variety has a singular line given by 
\begin{equation}
x_1x_2 = 0,\qquad x_3 = \xi_1^2x_1 + \xi_2^2x_2,\qquad x_4 = -(\eta_1^2x_2
+ \eta_2^2x_1).
\end{equation}
The case with $\xi_i\ne 0$, $\eta_1=0$, $\eta_2=0$ will be similar.
\item $\eta_1, \eta_2\neq 0$, $\eta_3=0$. There is no singularity.
\item \label{spbr3} $\eta_1=0$, $\eta_2,\eta_3\neq 0$. 
This has a line-singularity given by
\begin{equation}
\begin{split}
x_2 = \frac{\eta_2\xi_1}{\eta_3},\qquad x_4&=-\frac{\eta_2\eta_3}{\xi_1}, \\
\eta_2(x_1x_3 - \xi_2^2) - \eta_3\xi_1x_1 &=0, 
\end{split}
\end{equation}
\item 
$\eta_1=\eta_2=0$, $\eta_3\neq 0$. This has a singularity along $x_3$,
with $x_1=x_2=x_4=0$.
\item $\eta_1\neq 0$, $\eta_2$, $\eta_3=0$. This has a singular plane
along the solutions of 
\begin{equation}
x_1x_4 + \eta_1^2 = 0, \qquad x_2x_3 - \xi_1^2 = 0.
\end{equation}
\end{enumerate}
\item Finally, the cases with all the three $\xi_i$ turned on. 
When all $\eta_i\neq 0$, this has a line singularity
\begin{equation}\label{spbr2}
\begin{split}
x_1 &= \frac{\xi_2\xi_3}{\xi_1}\left[ 1+ \frac{(a-b) (c-a)}{x+bc}\right],\\
x_2 &= \frac{\xi_3\xi_1}{\xi_2}\left[ 1+ \frac{(a-b) (b-c)}{x+ca}\right],\\
x_3 &= \frac{\xi_1\xi_2}{\xi_3}\left[ 1+ \frac{(b-c) (c-a)}{x+ab}\right],\\
x_4 &= \frac{x}{\xi_1\xi_2\xi_3},\\
t &= 0,
\end{split}
\end{equation}
where we have defined $a=\eta_1\xi_1$, $b=\eta_2\xi_2$ and $c=\eta_3\xi_3$
and $x$ is an arbitrary complex number.

The different cases with some of the $\eta_i$ set to zero 
can be obtained from \eq{spbr2} by setting the parameters $a,b,c$ to 
zero accordingly. 
\end{enumerate}
Moreover, in the above discussion, we have assumed generic values 
of the six parameters $\xi_i$ and $\eta_i$, whenever non-zero. Several other 
special branches can be obtained by relating these non-vanishing parameters.
These can be obtained from the corresponding cases in the above list.
\subsection{The moduli space for other representations of $G$}
So far we have used the eight-dimensional regular representation of $G$,
constructed by tensoring the four-dimensional projective representation 
$\diag{\sigma_I, - \sigma_I}$, from \eq{REP} with $\id{M}$, with $M=2$.  
This corresponds to a single brane at the orbifold singularity. 
The theory with $N$ branes at the singularity can be obtained, 
in a similar way as above, 
using $\id{M}$, such that $4M=2^3N$. The moduli spaces for 
configurations with $N>1$ branes, or
equivalently $M>2$, can be found in the same manner as discussed 
above. However, it has been pointed 
out \cite{dt1,dt2} that string theory allows also the case with $M=1$. 
Let us discuss this case briefly in this subsection.

The quantities $z_I$ and $w_I$ appearing in \eq{higgs} are now 
numbers, instead of $2\times 2$ matrices.
The $M=1$ moduli space ensues by imposing the F-flatness conditions 
\eq{f-flatness} on the complex numbers $z_I$ and $w_I$. 

We can define the gauge-invariant polynomials as above. However, in the 
present case, the polynomials $\zpol{1234}{4}$ and $\wpol{1234}{4}$ are not
independent, but determined by polynomials of order two, \eg 
$2z_1w_2z_3w_4= \frac{1}{2}(a-b+3c)+ \xi_3z_3w_4+\eta_3z_1w_2$. 
Thus, we define only the four variables 
\begin{equation}
x_1 = z_I w_I, \quad I=1,2,3,4,
\end{equation}
as above. 

When $\eta_i=\xi_i=0$, we have, the F-flatness equations \eq{f-flatness},
can be partially solved with, \eg $z_1=z_2=w_1=w_2=0$. This leaves us 
with the following F-flatness condition
\begin{equation}\label{monom}
z_3w_4-z_4w_3=0,
\end{equation}
and the D-flateness equation, now reduced to
\begin{equation}
|z_3|^2+|z_4|^2 - |w_3|^2 - |w_4|^2 = 0.
\end{equation}
This space admits a toric description, thanks to the monomial relation
\eq{monom}. Following the steps outlined in Example 1 above, it can be 
seen that the moduli space is a $\C^2/\Z_2$-plane
\begin{equation}
xz -y^2 =0,
\end{equation}
where $x,y,z$ 
are three complex numbers. 
The same exercise can be repeated 
with other pairs of $z_i$ and $w_i$, corresponding to other 
possible solutions of the F-flatness equations. Since there are six
such solutions, six $\C^2/\Z_2$-planes arise and the moduli space
is given as the union of these six planes. These six planes
can be interpreted as associated with the invariant subspaces
of the six group elements each of which contribute a deformation.
The fractional brane may be a D3-brane wrapped on a vanishing
two-sphere transverse to the plane. However a confirmmation
of this needs more involved analysis.

When $\eta_i=0$, and $\xi_i\neq 0$, three of the equations 
in \eq{f-flatness} involving $z_4$ and $w_4$ are solved by
$w_I = s z_I$, where $s$ is a constant. The three rest 
can be rewritten as 
\begin{equation}
\begin{split}
x_1x_2 &= \xi_3^2/4,\\
x_2x_3 &= \xi_1^2/4,\\
x_3x_1 &= \xi_2^2/4,
\end{split}
\end{equation}
while $x_4 = sz_4^2$.
The solution to these equations furnishes the moduli space, which 
is a line given by 
\begin{equation}
x_1 = \frac{\xi_2\xi_3}{2\xi_1},\quad
x_2 = \frac{\xi_3\xi_1}{2\xi_2},\quad
x_3 = \frac{\xi_1\xi_2}{2\xi_3},
\end{equation}
with $x_4$ arbitrary.

Finally, when all $\xi_i$ and $\eta_i$ are non-zero, the moduli space 
is described by equations involving $x_i$, $i=1,2,3$ in a similar,
but more complicated way, as above. These can again be solved  for $x_i$,
while $x_4$ is left arbitrary. Thus, the moduli space is again a line.
\section{Conclusion}\label{conclud}
To summarise, we have studied
a D1-brane on the four-dimensional orbifold singularity 
$\C^4/(\Z_2\times\Z_2\times\Z_2)$, with 
discrete torsion. The resulting moduli 
space in absence of any deformation is the singular orbifold 
given by the equation $t^2=x_1x_2x_3x_4$ in $\C^5$, as found
earlier \cite{ahn}. 
The deformations of the moduli space in the presence 
of discrete torsion correspond to 
perturbations of the superpotential of the corresponding 
$(0,2)$ SYM in two dimensions, and are constrained by 
consistency requirements from string theory on the orbifold.
The moduli arising in the twisted sector of string theory now
deform away certain singularities
of codimensions one and two, which in turn correspond to 
some subsets of $\C^4$ fixed by certain elements of the group 
$G=\Z_2\times\Z_2\times\Z_2$. This is in harmony with 
expectations from conformal field theoretic description.
But the singularity with codimension 
three turns out to be stable, as the twisted sector 
of the closed string theory fails to provide the 
modes required for its deformation. That the stable singularity is 
a line and not a point, unlike its three-dimensional counterpart
\cite{dt1,dt2}, is rooted in the peculiarity of discrete torsion. 
As mentioned in \S\ref{proj}, the discrete torsion $\alpha$, in spite 
of its deceptive general appearance in \eq{eqnalpha}, is actually between 
two $\Z_2$ factors out of the three in the quotienting group 
$\Z_2\times\Z_2\times\Z_2$. By a suitable change of basis,
we can  thus think of the discrete torsion as 
affecting only a $\Z_2\times\Z_2$ subgroup of $G$, acting on a
subset $\C^3\subset\C^4$. Apart from the details of further quotienting 
by a $\Z_2$, it is the node of this $\C^3/(\Z_2\times\Z_2)$ \cite{dt1,dt2}, 
that gives rise to the singular line.

Indeed, this aspect of discrete torsion has featured in studies of 
mirror symmetry on this orbifold.
Starting from the Calabi-Yau manifold obtained as the blown up 
$\T^8/(\Z_2\times\Z_2\times\Z_2)$, which is the compact version 
of our case at hand, one can T-dualise the type-II 
string theory on this manifold
along the $\T^4$-fibres, to obtain the same string theory on 
the deformed $\T^8/(\Z_2\times\Z_2\times\Z_2)$ singularity. 
The T-dualities along the four directions of the 
$\T^4$ administers the ``right dose" of discrete 
torsion as in here, such that the two 
theories are mirror-dual to each other.
This situation provides a non-trivial demonstration of 
mirror symmetry \cite{acharya}.  
Thus, the situation considered in 
the present article is expected to be mirror-dual 
to the moduli space obtained by considering
D-brane in absence of discrete torsion \cite{ahn}.
A comment is in order. In absence of discrete torsion, a D1-brane
on the orbifold $\C^4/(\Z_2\times\Z_2\times\Z_2)$ resolves the orbifold
singularity with seven \FI parameters \cite{ahn}. In view of the
fact that for this resolution, $h^{11}=6$, as we found in \S\ref{closed}, 
this signifies that the resolved moduli space is smooth but not 
Calabi-Yau \cite{mohri}. It is not clear at the moment
if one should invoke the 
mirror principle beyond Calabi-Yau varieties\cite{bat} to incorporate another
parameter of deformation in the present case also, superceding the
restrictions arising from the twisted sector of string theory.
If so, then this will call for a formulation of consistency conditions 
beyond the stringy ones studied here.

Another implication of this configuration is related to the fact
that discrete torsion can be simulated through an antisymmetric
B-field background
\cite{vafa}. On $\C^3/(\Z_2\times\Z_2)$, the B-field has a non-zero 
field strength supported at the singular point as necessitated by 
supersymmetry \cite{vafwit}.
In the present case, by an S-duality transformation,
we may change the D-string into a fundamental string (in Type--IIB) and the
background NS-NS B-field to an RR B-field. 
Thus, the present analysis can also be interpreted
as describing a fundamental string at an orbifold
in presence of a background RR B field which has a 
non-zero field strength only on a line singularity
at the classical level. 

Finally, in the case of  orbifold singularities there exist parallel
brane configurations that give rise to the theory of branes at the 
singularity.  In particular, one can also map the desingularisation
moduli of the singularity to the parameters of the brane configurations
\cite{lust,tatar}.
It would be interesting to understand the analogous brane 
configurations corresponding to present case 
and identify the presence as well as the absence
of the various deformation modes, which will provide another 
way of looking at the orbifolds with discrete torsion. 
\section*{Acknowledgement} We would like to thank D Jatkar
and A Sen for illuminating discussions.

\end{document}